\documentclass[preprint,showpacs,preprintnumbers,amsmath,amssymb]{revtex4}
\usepackage{graphicx}
\usepackage{epsfig}
\usepackage{bm}
\usepackage{amsfonts}
\usepackage{subfigure}
\usepackage{multirow}
\usepackage{float}
\usepackage{color}
\usepackage{xcolor}
\usepackage{ulem}
\usepackage{todonotes}
\usepackage{makecell}
\usepackage{enumitem}
\usepackage{float}
\usepackage{amsmath}
\usepackage{hyperref}
\hypersetup{colorlinks=True,citecolor=blue}
\usepackage{amsfonts}
\usepackage{amssymb}
\usepackage{caption}
\usepackage{float}
\usepackage{bm}
\usepackage{graphicx}
\usepackage{color}
\usepackage{verbatim}
\usepackage{subfigure}
\usepackage{multirow}
\def\thickhline{\noalign{\hrule height1.5pt}}

\usepackage{enumitem}
\usepackage{float}
\usepackage{amsmath}
\usepackage{hyperref}
\hypersetup{colorlinks=True,citecolor=blue}
\usepackage{amsfonts}
\usepackage{amssymb}
\usepackage{caption}
\usepackage{float}
\usepackage{bm}
\usepackage{graphicx}
\usepackage{color}
\usepackage{verbatim}
\usepackage{subfigure}
\usepackage{multirow}

\def\beq{\begin{equation}}
\def\eeq{\end{equation}}
\def\ber{\begin{eqnarray}}
\def\eer{\end{eqnarray}}
\def\benu{\begin{enumerate}}
\def\eenu{\end{enumerate}}

\def\sq{\lower.25ex\hbox{\large$\Box$}}
\def \lleq {\lower0.9ex\hbox{ $\buildrel < \over \sim$} ~}
\def \ggeq {\lower0.9ex\hbox{ $\buildrel > \over \sim$} ~}

\def\prl{{Phys.\@ Rev.\@ Lett.\ }}
\def\prd{{Phys.\@ Rev.\@ D\ }}

\def\plb {{Phys.\@ Lett.\@ B\ }}

\begin{document}

\title{\textbf{Breaking the degeneracy of non-canonical quartic inflation by reheating considerations}}

\author{Iraj Safaei\footnote{isafaei@kashanu.ac.ir} and Kayoomars Karami\footnote{kkarami@uok.ac.ir}}
\address{\small{Department of Physics, University of Kurdistan, Pasdaran Street, P.O. Box 66177-15175, Sanandaj, Iran}}

\begin{abstract}
Here, the quartic inflationary potential $V(\phi)=\frac{\lambda}{4}\phi^4$ within a non-canonical framework characterized by a power-law Lagrangian is investigated. We demonstrate that the predictions of this model align with the  Planck 2018 observational data. We explore how the predictions of the model depend on the non-canonical parameter $\alpha$ and the number of $e$-folds $N$. Notably, the sound speed, non-Gaussianity parameter, scalar spectral index, and tensor-to-scalar ratio are all affected by variations in $\alpha$. However, the scalar spectral index exhibits a degeneracy with respect to variation  in $\alpha$, which can be broken by incorporating reheating consideration.
By applying a combination of theoretical and observational constraints on $(r-n_{\rm s})$, non-Gaussianity, and reheating parameters, we find that  the duration of inflation is constrained to the range  $55 \leq N\leq55.7$ $e$-folds for $60 \leq \alpha \leq 130$.
Finally, we investigate relic gravitational waves and demonstrate that their energy density spectrum  falls within the sensitivity range of gravitational waves detectors for this constrained range of $e$-folds.
\end{abstract}
\maketitle
%
%
\section{Introduction}
The Big Bang theory of cosmology provides an explanation for structure and evolution of the universe. The theory of cosmic inflation was proposed to complement the Big Bang theory.
 Inflation is an unknown phenomenon during which the universe expands exponentially.
In Einstein's gravitation framework, a negative pressure is required to justify this accelerated expansion.
The quantum fields can be the source of negative pressure. The scalar field associated with inflation is usually called inflaton. The nature of the inflaton is still unknown.
 However, the exact causes of cosmic inflation have remained unknown. It has been suggested that the discovery of certain particles in fundamental particle physics can explain the cause of cosmic inflation \cite{Starobinsky:1980,Guth:1981,Linde:1982,Albrecht:1982,Linde:1990,Baumann:2009}. One of the predictions of inflationary cosmology is the generation of primordial scalar fluctuations. The growth of these fluctuations can lead to the formation of large scale structures. Additionally, the perturbations caused by scalar and tensor fluctuations during inflation can impact the cosmic microwave background (CMB) radiation. These effects are utilized to determine the scalar spectral index ($n_{\rm s}$) and the tensor-to-scalar ratio ($r$), which are two important observational parameters of inflation\cite{Liddle:2000,Baumann:2009}. These parameters are commonly used to test the validity of inflationary models. However, similar values of $n_{\rm s}$ or $r$ can be obtained by adjusting one or more free parameters in different models  \cite{Liddle:2000, BICEP2:2018}. This leads to a problem of indistinguishability among the models. Therefore, relying solely on observations of CMB radiation is insufficient to thoroughly investigate and differentiate between various inflationary models \cite{Kallosh:2013a, Kallosh:2013, akrami:2020}. Other observable parameters, such as the parameters of reheating epoch and relic gravitational waves (GWs), should be considered to effectively discern these models \cite{Grishchuk:1974, Starobinsky:1979, Starobinsky:1981}.

The universe has undergone various periods following inflation, each characterized by a specific equation of state parameter, including radiation-dominated and matter-dominated epochs. After the end of inflation and prior to the onset of the radiation-dominated period, the universe enters a reheating phase. Despite significant theoretical advancements in understanding the early universe, the physics of the reheating epoch remains poorly understood.
 At the end of inflation, the kinetic energy of the inflaton overcomes its potential energy. The inflaton then begins to oscillate around its potential minimum, gradually giving up its energy to other fields.
 However, if this process is slow, the reheating takes place perturbatively and the scalar field oscillates for a much longer time and gradually releases its energy in the form of radiation in the universe \cite{Baumann:2009,Kofman:1994,Kofman:1996,Gialamas:2020,Odintsov:2023}. In this scenario, the scalar field oscillates for an extended period, and the equation of state parameter of this oscillatory phase is determined by the form of the inflaton potential
 at its minimum value \cite{Baumann:2009}. Considering the reheating epoch can also be useful for verifying inflationary models.
 Descriptive parameters of the reheating period, such as the reheating duration ($N_{\rm re}$), reheating temperature ($T_{\rm re}$), and the equation of state parameter ($\omega_{\rm re}$), are usually analyzed for this epoch.

In \cite{Martin:2010}, Martin and Ringeval examined the constraints of CMB radiation on the reheating temperature. In \cite{LiangDai:2014}, Dai et al. investigated observational constraints on the inflationary power-law potentials and found that, the reheating constraints could impose more restrictions on this class of potentials too. In \cite{Eshaghi:2016}, Eshaghi et al. studied the $\alpha$-attractors inflationary models, including T-model and E-model. They calculated the predictions of these models for the scalar spectral index $n_{s}$ and the tensor-to-scalar ratio $r$, and then applied limits based on the observational data of Planck 2015. Furthermore, they determined a range for $\alpha$ based on  the reheating constraints. In \cite{Mishra:2021}, Mishra et al. showed that reheating parameters, along with the spectrum of relic GWs, can be used to distinguish between
canonical and non-canonical inflation,
and to break the degeneracies between two $\alpha$-attractors models with respect to their free parameters.

All mentioned in above motivate us to study the observational constraints on degeneracy of the quartic inflationary potential, $V(\phi)=\frac{\lambda}{4}\phi^4$, in non-canonical framework with power-law Lagrangian. To do so, in section \ref{sec2}  a brief formulation of the inflationary model with quartic potential  in the non-canonical framework is explained.  In section \ref{sec3}, the parameters of the reheating epoch for the model are examined. In section \ref{sec4},  the density of relic GWs predicted by the model is compared with the sensitivity curves of GWs detectors. Finally, in section \ref{sec5}, the conclusions of our work are presented.

\hspace*{1.5cm}


\section{Non-canonical inflation}\label{sec2}
We start this section with the following generic action
\begin{equation}\label{eq:action}
S=\int{\rm d}^{4}x \sqrt{-g} \left[\frac{R}{16\pi G}+{\cal L}(X,\phi)\right] ,
\end{equation}
where $g$ is determinant of the metric tensor $g_{\mu\nu}$ and $R$ is the Ricci scalar. Also the Lagrangian density $\mathcal{L}(X,\phi)$ is a function of the kinetic energy $X\equiv\frac{1}{2}g_{\mu\nu}\partial^\mu \phi \partial^\nu \phi$ and the scalar field $\phi$.
In this work, we consider the power-law form of the Lagrangian density as follows \cite{Panotopoulos-2007, Unnikrishnan:2012, Rezazadeh:2015, Mishra:2022,Heydari:2024a,Heydari:2024b,Heydari:2024c}
\begin{equation}\label{Lagrangian}
{\cal L}(X,\phi) = X\left(\frac{X}{M^{4}}\right)^{\alpha-1} - V(\phi) ,
\end{equation}
where $\alpha$ is a dimensionless non-canonical parameter, $M$ is a non-canonical mass scale parameter with the dimension of mass and $V\left(\phi\right)$ denotes the inflationary potential.
For the Lagrangian density (\ref{Lagrangian}), the energy density $\rho_{\phi}$ and pressure $p_{\phi}$ of the scalar field are obtained as
\begin{equation}
\rho _\phi \equiv\left ( \frac{\partial{\cal L} }{\partial X} \right )\left ( 2X \right )-{\cal L}=\left( {2\alpha - 1}
\right)X{\left( {\frac{X}{{{M^4}}}} \right)^{\alpha  - 1}} +
V(\phi) \label{eq:rho},
\end{equation}
\begin{equation}
{p_\phi } \equiv{\cal L}=X{\left( {\frac{X}{{{M^4}}}} \right)^{\alpha  - 1}} - V(\phi ) .\label{eq:Lp}
\end{equation}
Here, we consider a spatially flat Friedmann-Robertson-Walker (FRW) metric given by
\begin{equation}
\label{eq:FRW}
{\rm d} {s^2} = {\rm d} {t^2} - {a^2}(t)\left( {{\rm d} {x^2} +
{\rm d} {y^2} + {\rm d} {z^2}} \right) ,
\end{equation}
where $a(t)$ denotes the scale factor in terms of the cosmic time $t$.
By utilizing the FRW metric, the kinetic term can be defined as $X=\dot{\phi}^2/2$, where dot represents derivative with respect to cosmic time.
Varying the action (\ref{eq:action}) with respect to the metric (\ref{eq:FRW}) and using Eqs. (\ref{eq:rho}) and (\ref{eq:Lp}), the first and second Friedmann equations are obtained as follows
\begin{equation}
H^{2} =\dfrac{1}{3M_p^2} \rho_{\phi} = \frac{1}{3M_p^2}\left[\left(2\alpha-1\right)X\left(\frac{X}{M^{4}}\right)^{\alpha-1} + V(\phi)\right] , \label{eq: FR-eqn1}
\end{equation}
\begin{equation}
\dot{H} = -\frac{1}{2M_p^2} (\rho _{\phi} +p_{\phi} ) = -\frac{1}{M_p^2}\alpha X\left(\frac{X}{M^4}\right)^{\alpha -1},
\label{eq:FR-eqn2}
\end{equation}
where $H \equiv \dot{a}/a$ is the Hubble parameter, and $M_p \equiv 1 / {\sqrt{8\pi G}}$ is the reduced Planck mass.
Furthermore, the equation of motion for the scalar field $\phi$ is obtained by considering the Lagrangian (\ref{Lagrangian}) and minimizing the action (\ref{eq:action}) with respect to $\phi$, as
\begin{equation}
\label{eq:KG}
\ddot \phi  + \frac{{3H\dot \phi }}{{2\alpha  - 1}}+ \left( {\frac{V'(\phi)}{{\alpha (2\alpha  - 1)}}}
\right){\left( {\frac{{2{M^4}}}{{{{\dot \phi }^2}}}} \right)^{\alpha- 1}} = 0,
\end{equation}
where $({}')$ is the derivative with respect to $\phi$.
It should be noted that for $\alpha=1$, all of the above equations take their canonical form.
In the following, the Hubble slow-roll parameters are defined as
\begin{equation}
\label{eq:H slow roll parameters}
\epsilon\equiv -\frac{\dot{H}}{H^2} ~ , ~ \eta \equiv  - \frac{\ddot{\phi}}{H \dot{\phi}} = \epsilon -\frac{\dot{\epsilon }}{2H\epsilon }.
\end{equation}
Under the slow-roll approximation, i.e. $ (\epsilon, \eta) \ll 1 $, the first Friedmann equation  (\ref{eq: FR-eqn1})  takes the following form
\begin{equation}
\label{eq:FR1-SR}
3 M_p^{2}H^2\simeq V(\phi).
\end{equation}
Also by neglecting $\ddot{\phi}$ in Eq. (\ref{eq:KG}), and using Eq. (\ref{eq:FR1-SR}), the equation of motion (\ref{eq:KG}) is simplified as follows
\begin{equation}\label{eq:KG-SR}
\dot \phi  =  -
\theta \left[ {\left( {\frac{M_p}{{\sqrt 3 \alpha }}}
\right)\left( {\frac{{\theta {V}'(\phi )}}{{\sqrt {V(\phi )} }}}
\right){{\left( {2{M^4}} \right)}^{\alpha  - 1}}}
\right]^{\frac{1}{{2\alpha  - 1}}},
\end{equation}
wherein $\theta=+1$, if $V'(\phi)>0 $ and $\theta=-1$, if $V'(\phi)<0$.
In the slow roll regime $ (\epsilon, \eta) \ll 1 $, using Eqs. (\ref{eq:FR1-SR}) and (\ref{eq:KG-SR})  one can show that the slow roll parameters (\ref{eq:H slow roll parameters}) can be expressed in terms of the inflationary potential as follows \cite{Unnikrishnan:2012}
\begin{align}
&\epsilon \simeq  \left [ \frac{1}{\alpha } \left ( \frac{3M^{4}}{V\left( \phi\right)} \right )^{\alpha -1}\left ( \frac{M_{p}{V}'\left( \phi\right) }{\sqrt{2}V\left(\phi\right)} \right )^{2\alpha }\right ]^{\frac{1}{2\alpha -1}} , \label{eq:epsilon V}\\
&\eta \simeq   \left ( \frac{\alpha \epsilon }{2\alpha -1} \right )\left ( 2\frac{V\left ( \phi  \right ){V}''\left ( \phi  \right )}{{V}'\left ( \phi  \right )^2} -1 \right ). \label{eq:eta V}
\end{align}
It is notable that, for $\alpha = 1$, Eqs. (\ref{eq:epsilon V}) and (\ref{eq:eta V}) reduce to $\epsilon \simeq \epsilon_V$ and $\eta \simeq \eta_V - \epsilon_V$, where $\epsilon_V \equiv \frac{M_p^2}{2}\left(\frac{V'}{V}\right)^2$ and $\eta_V \equiv M_p^2\left(\frac{V''}{V}\right)$  represent the canonical forms of the first and second potential slow-roll parameters, respectively.

As for the dynamics of scalar and tensor perturbations under the slow-roll approximation within the non-canonical framework, following \cite{Garriga:1999}, the power spectrum of scalar perturbation is given by
\begin{equation}\label{eq:Ps-SR}
{\cal P}_{s}=\frac{H^2}{8 \pi ^{2}M_p^{2}c_{s} \epsilon}\Big|_{c_{s}k=aH}.
\end{equation}
Here,  $c_{_s}$ is the sound speed of the scalar perturbation defined as
\begin{equation}
\label{eq:cs-NC} c_s^2 \equiv \frac{{\partial {p_\phi }/\partial X}}{{\partial {\rho _\phi }/\partial X}} = \frac{{\partial {\cal L}(X,\phi )/\partial X}}{{\left( {2X} \right){\partial ^2}{\cal L}(X,\phi )/\partial {X^2} + \partial {\cal L}(X,\phi )/\partial X}}.
\end{equation}
The amplitude of the scalar power spectrum at the pivot scale ($k_{*}=0.05~\text{Mpc}^{-1}$) has been measured by Planck observations as ${\cal P}_{s} (k_{\ast })=2.1\times 10^{-9}$ \cite{akrami:2020}.
Considering the power-law non-canonical Lagrangian (\ref{Lagrangian}), the square of the  sound speed (\ref{eq:cs-NC}) is simplified as follows
\begin{equation}\label{eq:cs2}
c_{s}^{2}=\frac{1}{2\alpha -1}.
\end{equation}
In order to prevent the classical and phantom instabilities, it is necessary that $0\le c_{s}^{2}\le 1$ and consequently from Eq. (\ref{eq:cs2}) we have $\alpha\ge 1$.
In the following, the scalar spectral index $n_{s}$ is obtained from the scalar power spectrum in the non-canonical framework and it can be written in terms of slow-roll parameters as follows
\begin{equation}
n_s - 1\equiv \frac{\mathrm{d} \ln {\cal P}_ {s}}{\mathrm{d} \ln k}=-4\epsilon+2\eta. \label{eq:inf_ns_SR}
\end{equation}
The value of the scalar spectral index $n_{\rm s}$ is constrained by  Planck 2018 TT, TE, EE + LowE + Lensing + BK18 + BAO at 68\% and 95\% CL  as \cite{bk18,Paoletti:2022, akrami:2018}
\begin{equation}\label{eq:ns95}
n_s = 0.9653_{-\,0.0041\,-\,0.0083}^{+\,0.0041\,+\,0.0107} .
\end{equation}
Similar to  scalar perturbations, the power spectrum of tensor perturbations is obtained as follows
\cite{Garriga:1999}
\begin{equation}\label{eq:Pt-SR}
{\cal P}_{t}=\frac{2H^2}{\pi ^{2}M_p^2}\Big|_{k=aH},
\end{equation}
and the tensor spectral index is calculated in terms of first slow-roll parameter as
\begin{equation}
\label{eq:ntt}
n_{t}\equiv \frac{\mathrm{d} \ln {\cal P}_{t}}{\mathrm{d} \ln k}= -2 \epsilon.
\end{equation}
Also from Eqs. (\ref{eq:Ps-SR}) and (\ref{eq:Pt-SR}), the tensor-to-scalar ratio is computed as follows
\begin{equation}
\label{eq:r}
r\equiv\frac{{\cal P}_t}{{\cal P}_{s}}= 16 c_s \epsilon .
\end{equation}
The upper limit on the tensor-to-scalar ratio from measurements of Planck 2018 TT, TE, EE + LowE + Lensing + BK18 + BAO at the 95\% CL is \cite{Paoletti:2022,bk18}
\begin{equation}\label{eq:r95}
r \leq 0.036.
\end{equation}
Combining Eqs. (\ref{eq:ntt}) and (\ref{eq:r}) results in consistency relation as
\begin{equation}\label{eq:r-nt}
r =-8 c_s n_t.
\end{equation}
In what follows, we consider the power-law potential given by
\begin{equation}\label{eq:plow}
V\left( \phi\right) =V_{0}\phi^n ,
\end{equation}
wherein $V_0$ is a constant with dimension of $M_p^{4-n}$.
Setting $\epsilon =1$ in Eq. (\ref{eq:epsilon V}), the scalar field value at the end of the inflation $\phi_{e}$ is obtained as follows
\begin{equation}\label{eq:phi e}
\phi_{e}=M_{p}\left[  \left( \frac{\mu^{4\left( \alpha -1\right)  }}{\alpha}\right)  \left(\frac{3M_{p}^{4-n}}{V_{0}} \right)  \left( \frac{n}{\sqrt{2}}\right) ^{2\alpha} \right] ^{\frac{1}{\gamma\left( 2\alpha -1\right)  }} ,
\end{equation}
where
\begin{equation}\label{eq:gamma}
\gamma \equiv \frac{2\alpha +n\left ( \alpha -1 \right )}{2\alpha -1}~,~~\mu \equiv\frac{M}{M_p}.
\end{equation}
The inflationary $e$-folds number from the end of inflation to the  CMB horizon crossing moment  is given by
\begin{equation}\label{eq:N1}
N=-\int_{\phi _e}^{\phi }\left ( \frac{H}{ \dot{\phi }} \right )\mathrm{d} \phi .
\end{equation}
By substituting Eqs. (\ref{eq:KG-SR}) and (\ref{eq:phi e}) into Eq. (\ref{eq:N1}) and performing some algebraic calculations, the following equation can be obtained
\begin{equation}\label{eq:N}
\phi (N)=M_p\left[ \left ( \frac{n\mu ^{4(\alpha -1)}}{\alpha } \right )^{\frac{1}{2\alpha -1}}\left ( \frac{6M_{p}^{4-n}}{V_0} \right )^{\frac{\alpha -1}{2\alpha -1}}\left ( N\gamma +\frac{n}{2} \right )\right]^{1/\gamma}.
\end{equation}
In the following, using Eqs. (\ref{eq:FR1-SR}) and (\ref{eq:epsilon V}), the power spectrum of scalar perturbations (\ref{eq:Ps-SR}) for the power-law potential (\ref{eq:plow}) is given by
\begin{equation}\label{eq:pk}
{\cal P}_s(k)=\left ( \frac{1}{72\pi ^2 c_s} \right )\left [ \left ( \frac{\alpha 6^\alpha }{n^{2\alpha }\mu ^{4(\alpha -1)}} \right )\left ( \frac{V_0}{M_p^{4-n}} \right )^{3\alpha -2} \right ]^{\frac{1}{2\alpha -1}}\left ( \frac{\phi }{M_p} \right )_{c_s k=aH}^{\gamma +n}.
\end{equation}
Replacing Eq. (\ref{eq:N}) into (\ref{eq:epsilon V}), (\ref{eq:eta V}) and using Eq. (\ref{eq:inf_ns_SR}), the scalar spectral index for the power-law potential (\ref{eq:plow}) in non-canonical framework with power-law Lagrangian reads
\begin{equation}\label{eq:phi n}
n_{s}=1-2\left( \frac{\gamma +n}{2N\gamma +n}\right).
\end{equation}
Moreover, the power spectrum of the tensor perturbations,  from Eq. (\ref{eq:Pt-SR}) and using Eq. (\ref{eq:FR1-SR}) for power-law potential (\ref{eq:plow}), is given by
\begin{equation}\label{eq:pt}
{\cal P}_{t}= \frac{2V_0}{3\pi^2M_p^{4-n}}\left ( \frac{\phi }{M_p} \right )_{ k=aH}^{n} .
\end{equation}
Thereafter, inserting Eq. (\ref{eq:N}) in (\ref{eq:epsilon V}) and using Eq. (\ref{eq:ntt}), the tensor spectral index for the power-law potential (\ref{eq:plow}) is calculated as follows
\begin{equation}\label{eq:nt1}
n_{t}=- \frac{2n}{2N\gamma+n} .
\end{equation}
Furthermore,  by inserting Eqs. (\ref{eq:pk}) and (\ref{eq:pt}) into Eq. (\ref {eq:r}) and using  (\ref{eq:N}), the tensor-to-scalar ratio is obtained  as follows
\begin{equation}\label{rn}
r=\frac{1}{\sqrt{2\alpha -1}}\left ( \frac{16n}{2N\gamma +n} \right ) .
\end{equation}
The $V_0$ parameter of the potential is obtained by inserting Eq.  (\ref{eq:N}) in  (\ref{eq:pk}) and using the observational constraint ${\cal P}_{s} (k_{\ast })=2.1\times 10^{-9}$ on the amplitude of the scalar power spectrum at the pivot scale ($k_{*}=0.05~{\rm Mpc}^{-1}$)  \cite{akrami:2020} as follows
\begin{equation}\label{eq:V0}
\frac{V_0}{M_p^{4-n}}=\left [ \frac{12 \pi^2 n {\cal P}_s(k_\ast )}{\sqrt{2\alpha -1}}\left ( \frac{\alpha }{n}\left ( \frac{1}{6\mu ^4} \right )^{\alpha -1} \right )^{\frac{n}{\gamma (2\alpha -1)}} \left ( \frac{2}{2N \gamma +n} \right )^{\frac{\gamma +n}{\gamma }}\right ]^{\frac{\gamma (2\alpha -1)}{2\alpha }}.
\end{equation}

\begin{figure}[H]
\begin{minipage}[b]{1\textwidth}
\vspace{-.1cm}
\centering
\subfigure[\label{phiN}]{\includegraphics[width=.48\textwidth]%
{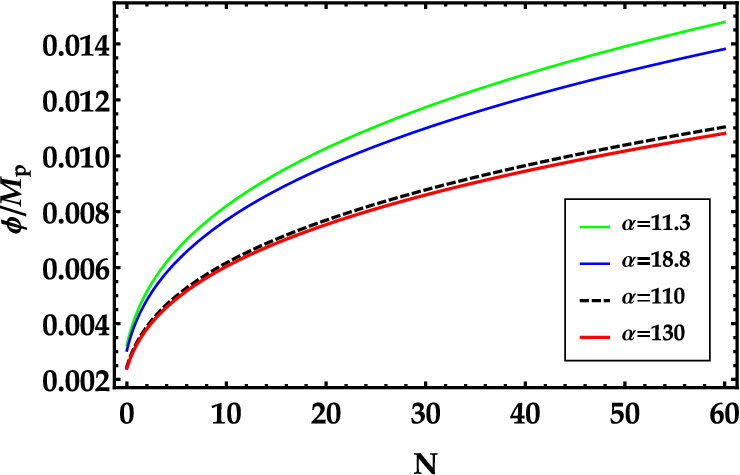}} \hspace{.1cm}
\centering
\subfigure[\label{HN}]{\includegraphics[width=.46\textwidth]%
{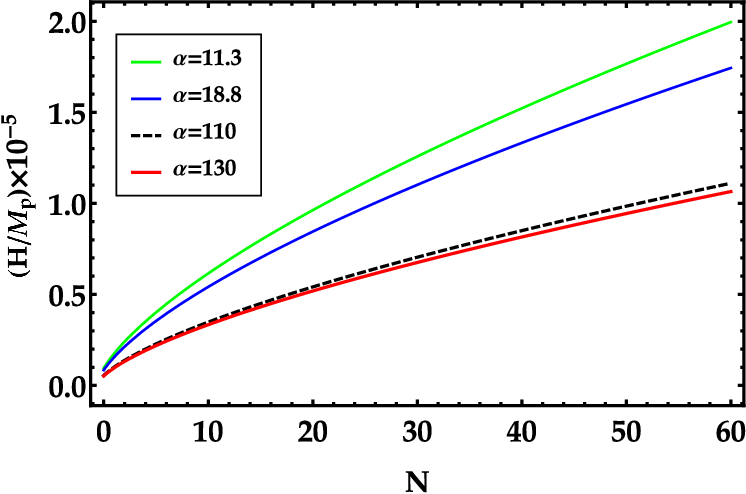}} \hspace{.1cm}
\hspace{.1cm}
\end{minipage}
\caption{Evolutions of (a) the scalar field $\phi$ and (b) and the Hubble parameter $H$,  for different values of $\alpha$. The green, blue, black dashed, and red lines represent $\alpha = 11.3, 18.8, 110$ and $130$, respectively.
 The end of inflation is set at $N=0$.}
\label{fig:back}
\end{figure}

Up to now, we reviewed the  basic analytical  equations for power-law potential (\ref{eq:plow}) in non-canonical framework with power-law Lagrangian (\ref{Lagrangian}). From now on, we will specifically focus on the quartic potential ($n=4$) as
\begin{equation}\label{eq:Higgs}
V\left ( \phi  \right )=\frac{\lambda }{4}\phi ^{4} ,
\end{equation}
where $\lambda=0.1$ is dimensionless self-interaction parameter \cite{Tanabashi:2018,Workman:2022}.
For the quartic potential (\ref{eq:Higgs}), with the help of Eqs. (\ref{eq:FR1-SR}) and (\ref{eq:N}), in Fig. \ref{fig:back} we plot the variations of the background quantities including the scalar field $\phi$ and the Hubble parameter $H$ as functions of the $e$-folds number $N$ for different values of $\alpha$. The figure shows that both $\phi$ and $H$ decrease during inflation, and that increasing $\alpha$ results in lower values of both quantities at a given $N$. This behaviour indicates that larger values of $\alpha$ correspond to a lower inflationary energy scale. It should be noted that the end of inflation is defined at $N=0$.

\begin{figure}[H]
\centering
\vspace{-0.2cm}
\scalebox{0.95}[0.95]{\includegraphics{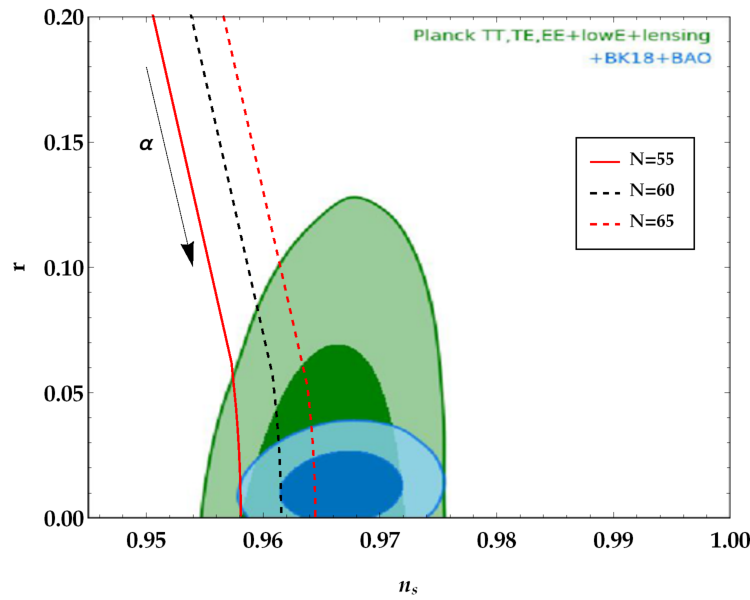}}
\vspace{-0.1cm}
\caption{The tensor-to-scalar ratio $r$ versus the scalar spectral index $n_{\rm s}$ with varying $\alpha$ parameter and for different $e$-folds number $N=$ 55 (solid red curve),
60 (dashed black curve), and 65 (dashed red curve).
Dark (light) green area represents the 68\% (95\%) CL of Planck 2018 TT, TE, EE+low E+lensing and dark (light) blue one shows the 68\% (95\%) CL of Planck 2018 TT, TE, EE+low E+lensing+BK18+BAO in the background.}
\label{fig:r-ns}
\end{figure}

Note that our motivation for considering the potential (\ref{eq:Higgs}) comes back to this fact that the prediction of this potential in the standard model is completely ruled out by the Planck observations \cite{bk18,Paoletti:2022, akrami:2018}. But in what follows, we will show that within the framework of non-canonical inflation for the special ranges of $\alpha$ parameter, the prediction of quartic potential can lie inside the allowed regions of the Planck 2018 data (see the $r-n_{\rm s}$ panel in Fig. \ref{fig:r-ns}).
For the quartic potential (\ref{eq:Higgs}), $V_0=\lambda/4$ and $n=4$, the scalar spectral index  (\ref{eq:phi n}) is simplified as follows
\begin{equation}\label{eq:ns4}
n_{s}=1-\left( \frac{\gamma +4}{N\gamma +2}\right),
\end{equation}
where $\gamma=\left(\frac{6\alpha-4}{2\alpha-1}\right)$ and the tensor-to-scalar ratio (\ref{rn}) reduces to
\begin{equation}\label{eq:r4}
r=\frac{1}{\sqrt{2\alpha -1}}\left ( \frac{32}{N\gamma +2} \right ) .
\end{equation}
Also for this potential by inserting $V_0=\lambda/4$ in Eq. (\ref{eq:V0}), the parameter $\mu $ is obtained as follows
\begin{equation}\label{eq:mu1}
\mu = 6 ^{-\frac{1}{4}}\left ( \frac{4}{\alpha } \right ) ^{\frac{-1}{4\left ( \alpha -1 \right )}} \left [ \left ( \frac{\sqrt{2\alpha -1}}{48\ \pi ^2 P_s\left ( k_{*}  \right )}\left ( \frac{\lambda }{4} \right )^{\frac{\alpha }{3\alpha -2}} \right )^{\frac{-1}{4 \left ( \alpha -1 \right )}}\left ( \left ( \frac{1}{N\gamma +2} \right )^{-\frac{\gamma +4}{\gamma }} \right )^{\frac{2-3\alpha }{8\left ( \alpha -1 \right )}} \right ].
\end{equation}

In addition to the scalar spectral index and tensor-to-scalar ratio, the non-Gaussianity parameter is another important observational parameter that can impose a strong constraint on proposed inflationary models. Notably, in non-canonical models, only the equilateral shape of non-Gaussianity parameter can be considered \cite{Baumann:2009}, which has been specified for our model  as follows \cite{Rezazadeh:2015, Chen:2007}
\begin{equation}\label{eq:fnl}
f_{\rm NL}^{\rm equil}=-\frac{275}{972}\left ( \frac{1}{c_{s}^{2}}-1 \right )=-\frac{275}{486}\left ( \alpha -1 \right ) .
\end{equation}
The Planck measurements determine the value of the equilateral non-Gaussianity parameter as \cite{akrami:2020}
\begin{equation}\label{eq:fnlPlanck}
f_{\rm NL}^{\rm equil}=-26\pm 47 .
\end{equation}
Regarding Eqs. (\ref{eq:fnl}) and (\ref{eq:fnlPlanck}), the value of the $\alpha$ parameter in the non-canonical framework should be $\alpha \leq 130$.

\begin{figure}[H]
\centering
\vspace{-0.2cm}
\scalebox{1}[1.1]{\includegraphics{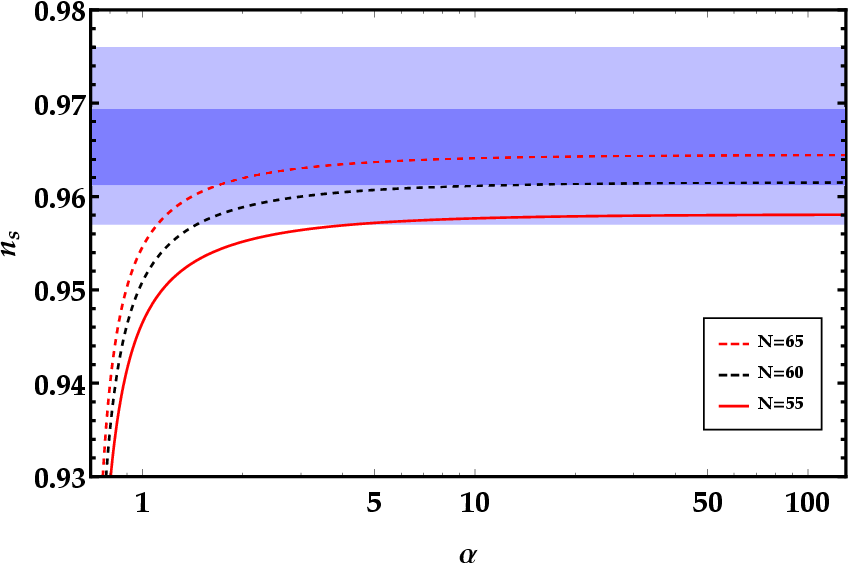}}
\vspace{-0.1cm}
\caption{The scalar spectral index  $n_{\rm s}$ versus non-canonical parameter $\alpha$ for different $e$-folds number
$N=55$ (solid red curve), $N=60$ (dashed black curve), $N=65$ (dashed red curve).
The dark and light blue regions show, respectively, the 68\% and 95\% CL of Planck 2018 TT, TE, EE + LowE + Lensing + BK18 + BAO data \cite{bk18,Paoletti:2022, akrami:2018}.}
\label{fig:ns}
\end{figure}

In Fig. \ref{fig:ns}, using Eq. (\ref{eq:ns4}), the scalar spectral index $n_{\rm s}$ is plotted as a function of the non-canonical parameter $\alpha$ for various values of the inflationary $e$-folds number. The permissible range for the scalar spectral index $n_{\rm s}$ according to Planck 2018 data in 68\% (95\%) CL is depicted in dark (light)  blue in the background.
Our estimations indicate that for $N=55$, the prediction of the model falls within the 95\% CL when $\alpha\geq 5$ (see the  solid red line in Fig. \ref{fig:ns}). Note that, as the value of $\alpha$ increases, the value of $n_{\rm s}$ remains almost constant, as depicted in Fig. \ref{fig:ns}. If we set $N=60$ (dashed black line), the estimated value for $n_{\rm s}$ can be consistent with the 68\% and 95\% CL when $\alpha \geq 2$ and $\alpha \geq 10$, respectively. Similar to previous case, the degeneracy of the scalar spectral index with respect to $\alpha$ parameter occurs at higher values of $\alpha$. For a larger $e$-folds number, our model aligns with observations at smaller $\alpha$ values.  However, the degeneracy of model persists at larger $\alpha$ values. So making a correct analysis solely based on CMB observations is impossible.


%

\begin{figure}[H]
\centering
\vspace{-0.2cm}
\scalebox{1}[1.1]{\includegraphics{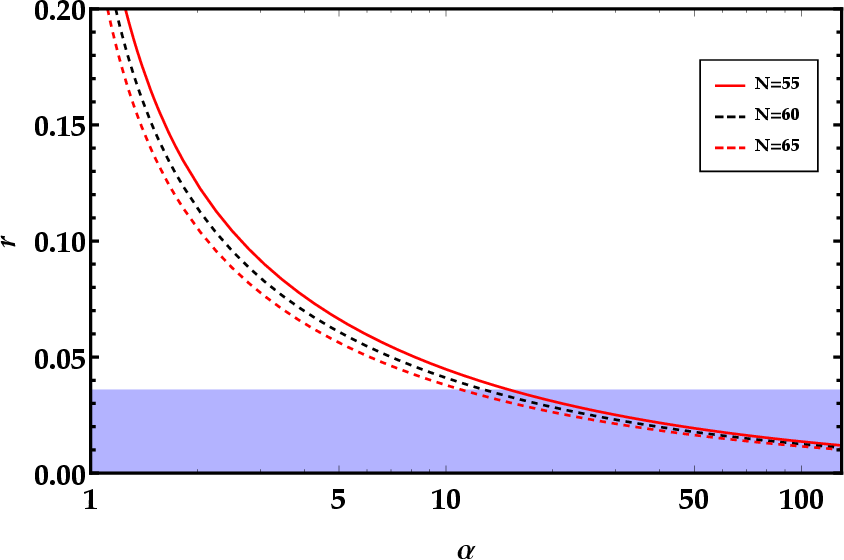}}
\vspace{-0.1cm}
\caption{The tensor-to-scalar ratio  $r$ versus the non-canonical parameter $\alpha$ for different $e$-folds number $N$. The light blue region shows $r\leq 0.036$ according to Planck 2018 TT, TE,
EE + LowE + Lensing + BK18 + BAO at 95\% CL \cite{bk18,Paoletti:2022, akrami:2018}.
}
\label{fig:r}
\end{figure}

In Fig. \ref{fig:r}, using Eq. (\ref{eq:r4}) the tensor-to-scalar ratio $r$ is depicted in terms of $\alpha$ parameter for various duration of inflationary era.
Furthermore, the allowed area for $r$ at 95\% CL according to the recent data of Planck's measurements (\ref{eq:r95}) is marked in light blue in this figure. As depicted in Fig. \ref{fig:r}, for $N=65$, the prediction of the model falls within the allowed region for $\alpha\geq 10$. Additionally, the cases associated with smaller $N$ can satisfy observational constraints at higher values of $\alpha$, as it can be seen from Fig. \ref{fig:r}.

Figure \ref{fig:r-ns} shows the $r-n_{\rm s}$ diagram in light of Planck and BICEP/Keck 2018 data (TT, TE, EE+lowE+lensing+BK18
+BAO) for different values of $N$ and varying $\alpha$. Note that on each curve, the parameter $\alpha$ increases from top to bottom.

In Table \ref{tab1}, the computed allowed ranges for the non-canonical parameters $\alpha$ and $\mu$, using  the constraints of $( r-n_{s})+f_{\rm NL}^{\rm equil}$ for different $e$-folds number $N$ are expressed. For instance, for the $e$-folds number $N=55$, the allowed range for the $\alpha$ parameter is $110\leq \alpha\leq 130$ at 95\% CL. Note that the $f_{\rm NL}^{\rm equil}$ parameter limits the maximum value of $\alpha$  to 130 in this model.

%

\hspace*{1.5cm}

\begin{table}[H]
\centering
\caption{
The allowed ranges for the model parameters $\alpha$ and $\mu \equiv M/M_p$ related to the permitted $e$-folds number $N$ in the non-canonical quartic inflationary model, considering a combination of various constraints from Planck and BICEP/Keck 2018 data on ($r-n_{\rm s}$), the equilateral non-Gaussianity parameter $f_{\rm NL}^{\rm equil}$, reheating $(\omega_{\rm re}+N_{\rm re}+T_{\rm re}$), and relic GWs. Here, the parameter $\mu $ is estimated from Eq. (\ref{eq:mu1}).
%
}
\label{tab1}
\resizebox{\columnwidth}{!}{%
\tiny
\begin{tabular}{c|cccc|cc}
\thickhline
\multirow{3}{*}{$N$} &
  \multicolumn{4}{c|}{\multirow{2}{*}{$ (r-n_{\rm s})+ f_{\rm NL}^{\rm equil}$}} &
  \multicolumn{2}{c}{\makecell[c]{$ (r-n_{\rm s})+ f_{\rm NL}^{\rm equil}$ \\ $+ \omega_{\rm re}+N_{\rm re}+T_{\rm re}$ }} \\ \cline{6-7}
 &
  \multicolumn{4}{c|}{} &
  \multicolumn{2}{c}{\makecell[c]{$ (r-n_{\rm s})+ f_{\rm NL}^{\rm equil}$ \\ $+ \omega_{\rm re}+N_{\rm re}+T_{\rm re}+{\rm GWs} $}} \\ \cline{2-7}
 &
  \multicolumn{1}{c|}{$\alpha (95\% \text{ CL})$} &
  \multicolumn{1}{c|}{$\alpha (68\% \text{ CL})$} &
  \multicolumn{1}{c|}{$\mu (\times10^{-5})~(95\% \text{ CL})$} &
  $\mu (\times10^{-5})~(68\% \text{ CL})$ &
  \multicolumn{1}{c|}{$\alpha (95\% \text{ CL})$} &
  $\mu (\times10^{-5})~(95\% \text{ CL})$ \\ \thickhline
$55$ &
  \multicolumn{1}{c|}{$[110,130]$} &
  \multicolumn{1}{c|}{$-$} &
  \multicolumn{1}{c|}{$[2.23,2.30]$} &
  \multicolumn{1}{c|}{$-$} &
  \multicolumn{1}{c|}{$[110,130]$} &
  $[2.23,2.30]$ \\ \hline
$55.3$ &
  \multicolumn{1}{c|}{$[80,130]$} &
  \multicolumn{1}{c|}{$-$} &
  \multicolumn{1}{c|}{$[2.22,2.40]$} &
   \multicolumn{1}{c|}{$-$}&
  \multicolumn{1}{c|}{$[80,130]$} &
  $[2.22,2.40]$ \\ \hline
$55.7$ &
  \multicolumn{1}{c|}{$[60,130]$} &
  \multicolumn{1}{c|}{$-$} &
  \multicolumn{1}{c|}{$[2.21,2.49]$} &
  \multicolumn{1}{c|}{$-$} &
  \multicolumn{1}{c|}{$[60,65]$} &
  $[2.46,2.49]$ \\ \hline
$60$ &
  \multicolumn{1}{c|}{$[18.8,130]$} &
  \multicolumn{1}{c|}{$-$} &
  \multicolumn{1}{c|}{$[2.07,2.59]$} &
  $-$ &
  \multicolumn{1}{c|}{$-$} &
  $-$ \\ \hline
$65$ &
  \multicolumn{1}{c|}{$[11.3,130]$} &
  \multicolumn{1}{c|}{$[25,130]$} &
  \multicolumn{1}{c|}{$[1.93,2.34]$} &
  $[1.93,2.39]$ &
  \multicolumn{1}{c|}{$-$} &
  $-$ \\ \thickhline
\end{tabular}%
}
\end{table}

\hspace*{1.5cm}

\section{Reheating epoch} \label{sec3}

As mentioned before, at the end of inflation, the kinetic energy of the inflaton overcomes its potential energy. Thence the inflaton oscillates around the minimum of the potential  and gradually releases its energy. This epoch is called reheating and its nature is still not fully understood \cite{Baumann:2009}. Considering the reheating epoch can also be useful for verifying inflationary models. In this epoch, the equation of state parameter ($\omega_{\rm re}$), the duration of the reheating epoch ($N_{\rm re}$), and the reheating temperature ($T_{\rm re}$) are usually investigated. The duration of reheating epoch and reheating temperature both depend on the equation of state parameter. The most important constraint that the duration of the reheating epoch imposes on the model is that its value must be positive, i.e., $N_{\rm re}\geq0$. Additionally, the duration of the reheating epoch should be significantly shorter than the duration of the inflation epoch and it can be obtained as follows \cite{Martin2015,Munoz2015,Cook2015,German2020}
\begin{equation}\label{eq:Nre}
N_{\rm re}=\left(\frac{4}{1-3 \omega _{\rm re}}\right)\left [ -\frac{1}{4}\ln \left ( \frac{30}{\pi ^2g_{\rm re}} \right )-\frac{1}{3} \ln \left ( \frac{11 g_{\rm re}}{43} \right )-\ln \left ( \frac{k_\ast }{a_0 T_0} \right )-\ln \left ( \frac{\rho_e^{1/4}}{H_{\ast}} \right )-N\right ] ,
\end{equation}
where  $g_{\rm re}=106.75$ is the effective number of relativistic degrees of freedom for energy during the reheating epoch. Also $k_\ast=0.05~{\rm Mpc^{-1}}$ is the pivot scale,  $a_0=1$ is the scale factor at the present time, $T_0=2.725~{\rm K}$ is the current temperature of CMB radiation,  $\rho _e\equiv \rho_{\phi_ e}$ is the energy density of the inflaton field at the end of inflation, $H_{\ast}$ is the Hubble parameter at the horizon crossing time and $N$ defines the duration of inflation from the end of inflation until  the horizon passing moment . Using the conversion factor $1~{\rm Mpc^{-1}}=6.39\times10^{-39}{\rm GeV}$, Eq. (\ref{eq:Nre}) can be simplified as follows
\begin{equation}\label{eq:Nre1}
N_{\rm re}=\frac{4}{1-3 \omega _{\rm re}}\left[61.6-\ln \left (\frac{\rho_e^{1/4}}{H_{\ast}} \right)-N\right],
\end{equation}
where the value of $H_{\ast}$ can be estimated from Eq. (\ref{eq:Ps-SR}) as follows
\begin{equation}\label{eq:Hi}
H_{\ast}=\sqrt{ 8 \pi ^2 M_p^2 c_s {\cal P}_s(k_\ast ) \epsilon }.
\end{equation}
Here, ${\cal P}_{\rm s}(k_\ast ) = 2.1\times 10^{-9}$ denotes the amplitude of the scalar power spectrum at the CMB pivot scale. It is known that, the equation of state parameter is defined as
\begin{equation}\label{eq:Omega total}
\omega_{\phi}\equiv \frac{p_\phi}{\rho_\phi}=-1-\frac{2}{3}\frac{\dot{H}}{H^2}=-1+\frac{2}{3}\epsilon.
\end{equation}
The value of the equation of state parameter at the end of inflation ($\epsilon=1$) is $\omega_{\phi_ e}=-\frac{1}{3}$, and consequently
\begin{equation}\label{eq:phi-rho1}
p_{e}=-\frac{1}{3}\rho_e,
\end{equation}
where $p_{e}\equiv p_{\phi_ e}$. By substituting Eqs. (\ref{eq:rho}) and (\ref{eq:Lp}) into Eq. (\ref{eq:phi-rho1}), the energy density at the end of inflation $\rho_e$ in the non-canonical framework with power-law Lagrangian (\ref{Lagrangian}) can be obtained by
\begin{equation}\label{eq:rho end}
\rho _e=\left(\frac{3\alpha }{\alpha +1}\right)V_e,
\end{equation}
where $V_e$ is the inflationary potential value  at the end of inflation.

For non-canonical action with the power-law Lagrangian (\ref{Lagrangian}), the equation of state parameter in the reheating epoch can be calculated as follows \cite{Unnikrishnan:2012}
\begin{equation}\label{action}
1+\left \langle \omega _{\phi } \right \rangle=\left ( \frac{2\alpha }{2\alpha -1} \right )\left [ \int_{0}^{\phi _{m}}{\mathrm{d}\phi \left ( 1-\frac{V\left (\phi  \right ) }{V\left (\phi_{m}  \right )} \right ) }^{\frac{2\alpha -1}{2\alpha }}\right ]\left [\int_{0}^{\phi _{m}}{\mathrm{d}\phi \left ( 1-\frac{V\left (\phi  \right ) }{V\left (\phi_{m}  \right )} \right ) }^{\frac{-1}{2\alpha }}  \right ]^{-1} ,
\end{equation}
where $\langle ~\rangle$ represents the average value over one oscillation cycle of the inflaton.

 Similar to the standard inflation in the canonical framework, we assume that the timescale of the scalar field oscillation is much smaller than that of the expansion of the universe. In this limit, the time variations of $\rho_{\phi}$ during each cycle is small enough to approximate $\rho_\phi \simeq V(\phi_m)$, where $\phi_m$ is the maximum value of the field during a given cycle. According to Eq. (\ref{action}), in the framework of non-canonical inflation with a power-law Lagrangian for the potential $V\left(\phi\right)=V_{0}\phi^{n}$, the equation of state parameter in the reheating epoch can be calculated as follows
\begin{equation}\label{eq:wren}
\omega _{\rm re}=\left \langle \omega _{\phi } \right \rangle=\frac{n-2\alpha }{n\left ( 2\alpha -1 \right )+2\alpha },
\end{equation}
which for $n=4$, it can be written as
\begin{equation}\label{eq:wre4}
\omega _{\rm re}=\frac{2-\alpha }{5\alpha -2} .
\end{equation}

\begin{figure}[H]
\centering
\includegraphics[width=.65\textwidth]{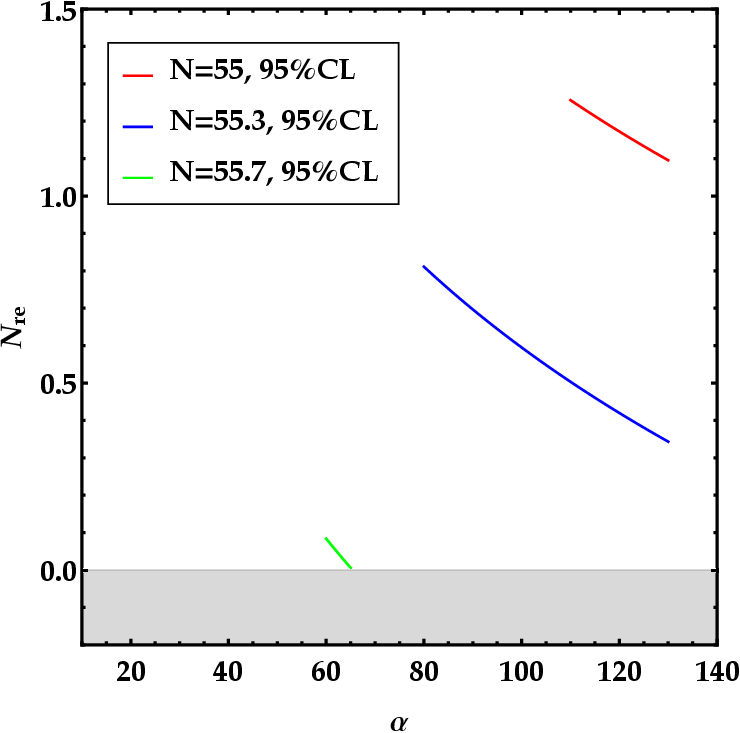}
\caption{
Duration of the reheating stage $N_{\rm re}$ versus the $\alpha$ parameter for different values of the $e$-folds number $N$ according to Table \ref{tab1}. The gray forbidden area corresponds to negative values of $N_{\rm re}$.}
\label{fig:Nre}
\end{figure}

In the following, we calculate the duration of the reheating epoch $N_{\rm re}$ for non-canonical action with power-law Lagrangian (\ref{Lagrangian}) and quartic potential (\ref{eq:Higgs}) by inserting Eqs.  (\ref{eq:Hi}), (\ref{eq:rho end}) and (\ref{eq:wre4}) in (\ref{eq:Nre1}).
%
%
In Fig. \ref{fig:Nre}, we plot the reheating duration $N_{\rm re}$ versus $\alpha$ for fixed values of the $e$-folds number $N$, as listed in Table \ref{tab1}.
Here, a clear dependence of $N_{\rm re}$ on $\alpha$ is evident.
Therefore, in the ($N_{\rm re} - \alpha$) plane, the degeneracy with respect to $\alpha$ can be broken, as different values of $\alpha$ yield distinguishable reheating durations.
It is worth noting that the gray area corresponds to negative values of $N_{\rm re}$, which are non-physical and therefore excluded.
In this regard, the condition $N_{\rm re}\geq 0$ imposes an upper bound on the duration of the inflationary epoch in the present model as $N=55.7$ and subsequently refine the upper bound on the non-canonical parameter to $\alpha=65$, as shown in Table \ref{tab1}.

\hspace*{1.5cm}

In addition to $N_{\rm re}$, the reheating temperature $T_{\rm re}$ can also be used to survey inflationary models and it can be calculated as follows \cite{Martin2015,Munoz2015,Cook2015,German2020}
\begin{equation}
T_{\rm re} = \left ( \frac{30 \rho_{\rm e}}{\pi ^{2}g_{\rm re}} \right )^{1/4}\exp\left[-\frac{3}{4}( 1+\omega_{\rm re})N_{\rm re}\right] .
\label{eq:CMB_reheat_Tre_SR}
\end{equation}
This parameter depends on  $\omega_{\rm re}$ and $N_{\rm re}$. Since the reheating stage occurs after the end of inflation and before the beginning of the Big Bang Nucleosynthesis (BBN) stage, the upper and lower limits on the reheating temperature are defined as follows \cite{Martin2015}
\beq
10 \, {\rm MeV} \leq T_{\rm re} \leq 5\times10^{15}\, {\rm GeV} .
\label{eq:bound_Tre}
\eeq

\begin{figure}[H]
\centering
\includegraphics[width=.65\textwidth]{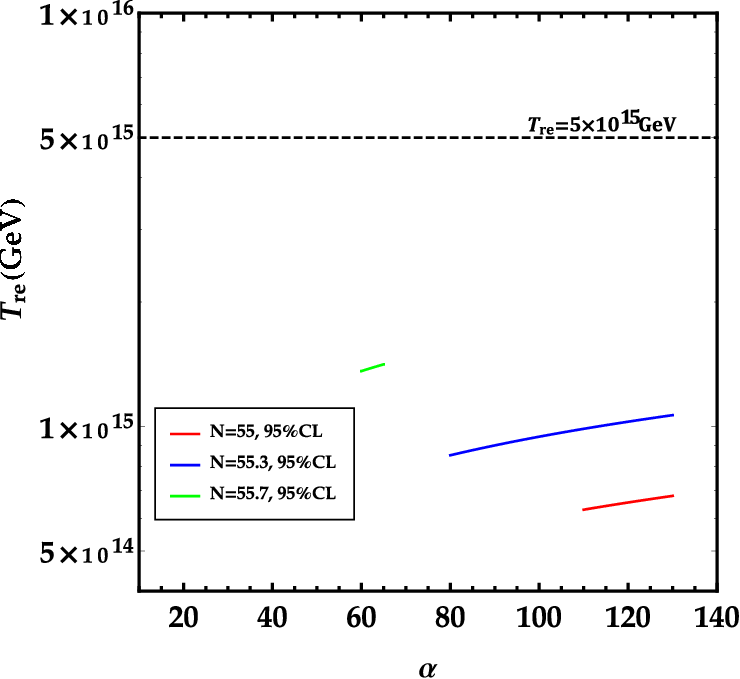}
\caption{
Reheating temperature $T_{\rm re}$ versus the parameter $\alpha$ for different values of the $e$-folds number $N$, as listed in Table~\ref{tab1}. The maximum allowable reheating temperature, $T_{\rm re} = 5 \times 10^{15}~\rm{GeV}$, is indicated by the horizontal dashed black line at the top.}
\label{fig:Tre}
\end{figure}

Figure \ref{fig:Tre} shows the variation of $T_{\rm re}$ as a function of $\alpha$ for different fixed values of the $e$-folds number $N$, according to Table \ref{tab1}.
The horizontal line at the top of this figure indicates the upper limit of the reheating temperature.
In this case, a clear dependence on $\alpha$ is visible, and thus, in the ($T_{\rm re}-\alpha$) plane, the degeneracy can be broken.


\section{The relic gravitational waves} \label{sec4}

Inflation theory predicts that, primordial tensor fluctuations of the inflaton field can lead to  production of gravitational waves (GWs). During inflation, these fluctuations extend beyond the Hubble horizon, and upon reentering, they generate relic GWs \cite{Starobinsky:1979, allen88, sahni:1990, sami2002, dany_18, dany_19, Bernal:2019lpc}.  Due to their weak interaction with matter and radiation, these relic GWs serve as a valuable tool for probing the early universe. The present frequency of the originated relic GWs from the cosmic inflation can be calculated in terms of temperature using the following equation \cite{sahni:1990,dany_19,sami2002}
\beq
f=7.36\times 10^{-8} {~\rm Hz} \left ( \frac{g_{0}^{s}}{g_{T}^{s}} \right )^{\frac{1}{3}}\left ( \frac{g_{T}}{90} \right )^{\frac{1}{2}}\left ( \frac{T}{\rm GeV} \right ) ,
\label{eq:GW_f_master}
\eeq
where $g_0^s=3.94 $ and $g_T^s=106.75 $ are the effective number of relativistic degrees of freedom in entropy at the present time and each epoch with temperature $T$, respectively. Furthermore,  $g_T=106.75$  is the effective number of relativistic degrees of freedom in energy. The allowed ranges for the reheating temperature $T_{\rm re}$ and the corresponding frequency of GWs in the reheating epoch $f_{\rm re}$ are listed in Table \ref{tab GW1} for several values of $\alpha$ parameter according to Table \ref{tab1}. The present density parameter spectrum of relic GWs in the radiation-dominated and reheating periods are calculated from the following two equations (for more information, see Ref. \cite{sahni:1990,dany_19,sami2002})
\ber
\mbox{{ Radiation Dominated:}} ~~   \Omega _{\rm GW_0}^{(\rm RD)}(f)=\left ( \frac{1}{24} \right )r~{\cal P}_s (k_{\ast })\left ( \frac{f}{f_{\ast }} \right )^{n_{t}}\Omega _{r_0}~,~f_{\rm eq}< f\leq f_{\rm re} ,\,
\label{eq:GWs_spectrum_2a}\\
\mbox{{reheating:}} ~~ \Omega _{\rm GW_0}^{(\rm re)}(f)  = \Omega _{\rm GW_0}^{(\rm RD)}(f) \left (\frac{f}{f_{\rm re}}\right )^{2\left (\frac{\omega_{\rm re}-1/3}{\omega_{\rm re}+1/3}\right )},~~ f_{\rm re} < f \leq f_{e}\, ,
\label{eq:GW_spectrum_2b}
\eer
where $\Omega_{r_0}=2.47\times10^{-5}h^{-2}$ is the radiation density parameter at present time, $f_*=2.4 \times 10^{-16}~({\rm Hz})$ is the frequency of GWs at the CMB pivot scale $k_\ast=0.05~{\rm Mpc^{-1}}$, $f_{\rm eq}=1.7\times 10^{-17}~({\rm Hz})$ represents the frequency of GWs at the matter-radiation equality and $f_e=4.3\times 10^{8}~({\rm Hz})$ is the frequency of GWs at the end of inflation.


\begin{table}[H]
  \centering
  \caption{The allowed values of reheating temperature $ T_{\rm re}$ and the corresponding frequency $f_{\rm re}$ of GWs in the reheating epoch computed by Eq. (\ref{eq:GW_f_master}) for several  values of $\alpha$. See also Fig. \ref{fig:GW}.
%
  }
\resizebox{.6\textwidth}{!}{\scriptsize
\begin{tabular}{cccc}
\thickhline
$N$\qquad  &
\qquad$\alpha$\qquad &
\qquad $T_{\rm re}/{\rm GeV} $\qquad &
\qquad $f_{\rm re}/{\rm Hz} $\qquad  \\
\thickhline
\quad  \multirow{2}{*}{55} \quad\qquad & 
\qquad$ 110 $ \qquad  &
\qquad $6.29\times 10^{14} $\qquad &
\qquad $ 2.29\times10^{6}$  \\
\quad &
\qquad$ 130 $ \qquad  &
\qquad $ 6.80\times10^{14}  $\qquad &
\qquad $ 1.81\times10^{7} $  \\
\hline
\quad  \multirow{2}{*}{55.3} \quad\qquad & 
\qquad$ 80 $ \qquad  &
\qquad $8.52\times 10^{14} $\qquad &
\qquad $ 2.27\times10^{7}$  \\
\quad &
\qquad$ 130 $ \qquad  &
\qquad $ 1.07\times10^{15}  $\qquad &
\qquad $ 2.84\times10^{7} $  \\
\hline
\quad  \multirow{2}{*}{55.7} \quad\qquad & 
\qquad$ 60 $ \qquad  &
\qquad $1.36\times 10^{15} $\qquad &
\qquad $ 3.63\times10^{7}$  \\
\quad &
\qquad$ 65 $ \qquad  &
\qquad $ 1.41\times10^{15}  $\qquad &
\qquad $ 3.77\times10^{7} $  \\
\hline
%

\thickhline
\end{tabular}}
 \label{tab GW1}
\end{table}

\begin{figure}[H]
\begin{minipage}[b]{1\textwidth}
\vspace{-.1cm}
\centering
\subfigure[\label{GW55}]{\includegraphics[width=.58\textwidth]%
{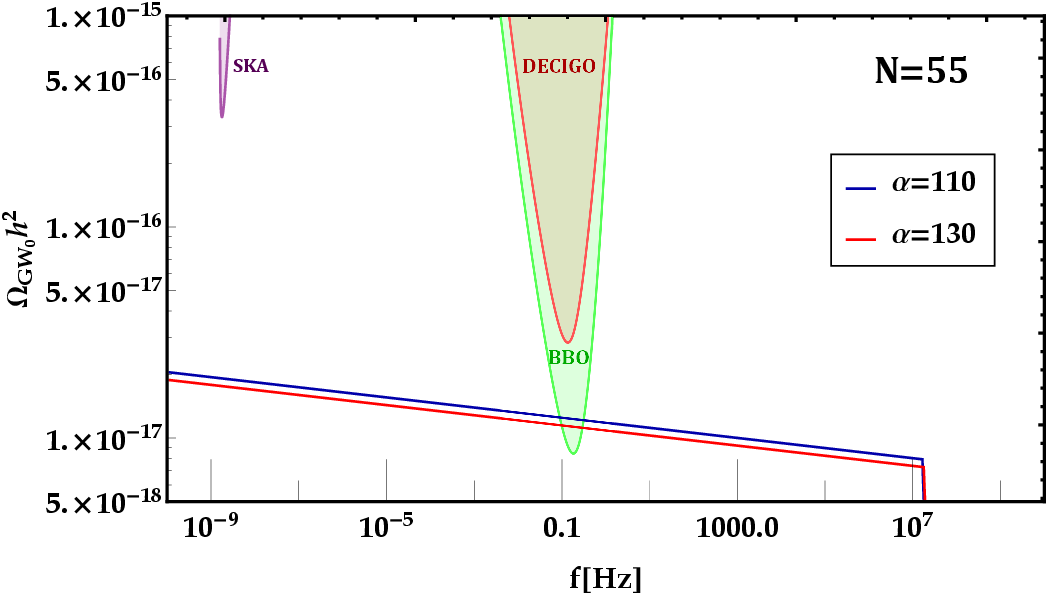}} \hspace{.1cm}
\centering
\subfigure[\label{GW553}]{\includegraphics[width=.58\textwidth]%
{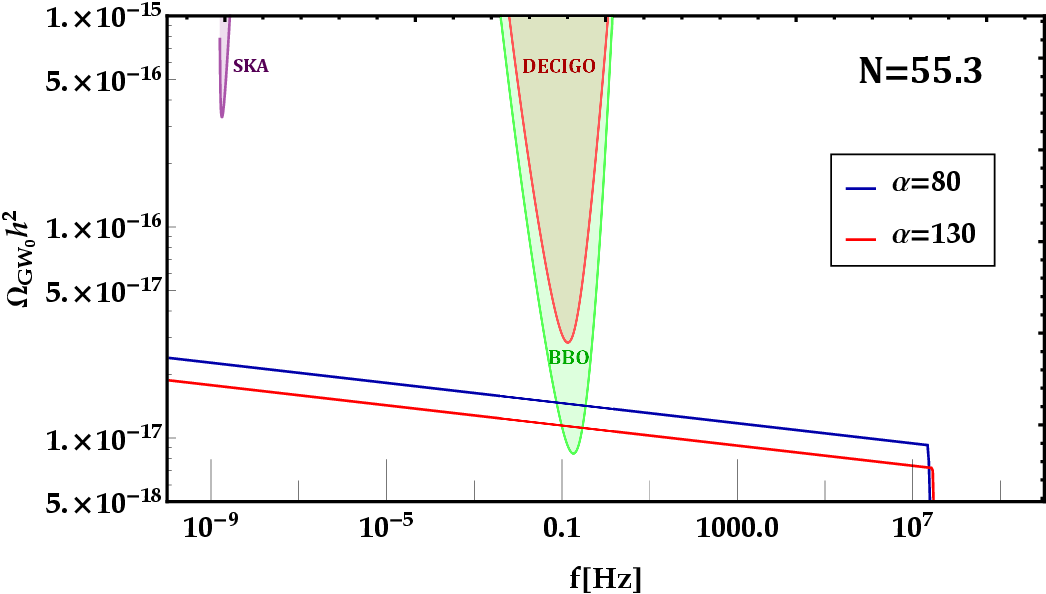}} \hspace{.1cm}
\centering
\subfigure[\label{GW557}]{\includegraphics[width=.58\textwidth]%
{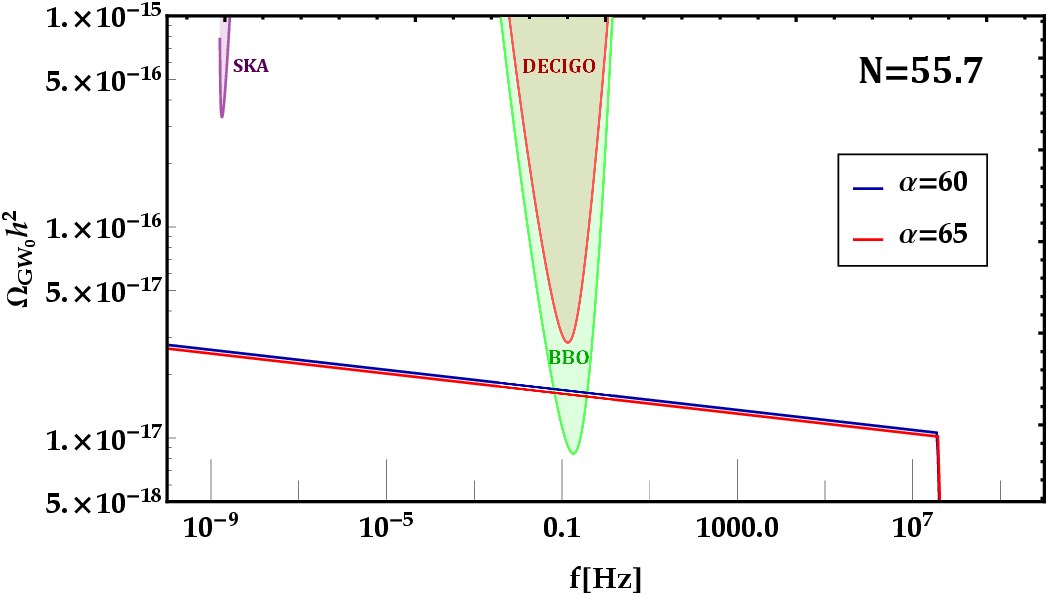}} \hspace{.1cm}
\end{minipage}
\vspace{-.4cm}
\caption{The energy density spectrum of relic GWs as a function of frequency for various $e$-folds numbers (a) $N=55$, (b) $N=55.3$, and (c) $N=55.7$, corresponding to different values of $\alpha$. The shaded regions represent the sensitivity ranges of  GW detectors, including
BBO \cite{Yagi:2011BBODECIGO,Yagi:2017BBODECIGO,Harry:2006BBO,Crowder:2005BBO,Corbin:2006BBO},
DECIGO \cite{Yagi:2011BBODECIGO,Yagi:2017BBODECIGO,Seto:2001DECIGO,Kawamura:2006DECIGO,Kawamura:2011DECIGO},
SKA \cite{ska,skaCarilli:2004,skaWeltman:2020}.
In each graph, the blue (red) curve  corresponds to the minimum (maximum) allowed value of $\alpha$, as given in Table \ref{tab1}.
The break in each spectrum marks the reheating frequency $f_{\rm re}$, which is related to the reheating temperature $T_{\rm re}$ listed in Table \ref{tab GW1}.
}
\label{fig:GW}
\end{figure}

Figure \ref{fig:GW} shows the present density parameter spectra of GWs for different values of the inflationary $e$-folds number $N$, calculated using Eqs.~(\ref{eq:GW_f_master})-(\ref{eq:GW_spectrum_2b}).
The green, red, and purple shaded regions represent the sensitivity ranges of the Big Bang Observer (BBO)~\cite{Yagi:2011BBODECIGO,Yagi:2017BBODECIGO,Harry:2006BBO,Crowder:2005BBO,Corbin:2006BBO}, the DECi-hertz Interferometer Gravitational-wave Observatory (DECIGO)~\cite{Yagi:2011BBODECIGO,Yagi:2017BBODECIGO,Seto:2001DECIGO,Kawamura:2006DECIGO,Kawamura:2011DECIGO}, and the Square Kilometre Array (SKA)~\cite{ska,skaCarilli:2004,skaWeltman:2020}, respectively.

The reheating temperatures and gravitational wave frequencies at the breakpoints of  the spectra are listed in Table~\ref{tab GW1}.
In each graph, the blue and red curves represent the present energy density of GWs, corresponding to the maximum and minimum allowed values of $\alpha$ for each $e$-folds number $N$, respectively.

It can be observed that, the curves fall within the sensitivity range of GWs detectors. Hence, with the advancement of these detectors, there is a possibility of detecting the predicted relic GWs by this model. Moreover, varying in $\alpha$ parameter leads to distinct values of relic GWs density.


\section{Conclusion}\label{sec5}

In this work, we investigated the quartic potential in non-canonical inflationary model with power-law Lagrangian (\ref{Lagrangian}). We showed that, in contrary to the standard inflation, the predictions of this potential in non-canonical framework are consistent with the observational data of Planck and BICEP/
Keck 2018. Among the free parameters of the model, only the changes of $\alpha$ parameter have a significant effect on the predictions of the model.

We checked the model based on the changes of $\alpha$ considering different values for the duration of the inflation epoch $N$. To determine the allowed range of $\alpha$ parameter, we used theoretical considerations as well as observational constraints. We showed that the sound speed parameter $c_s$, equilateral non-Gaussianity $f_{\rm NL}^{\rm equil}$, scalar spectral index $n_{\rm s}$, and tensor-to-scalar ratio $r$ bound the parameter $\alpha$ (see Table \ref{tab1}). The minimum value of $\alpha$ was determined to be larger than one based on the constraint on $c_s$. The equilateral non-Gaussianity constraint also limits the maximum value of $\alpha$ to $\alpha\leq 130$. The $r-n_{\rm s}$ curve sets the lower bound on $\alpha$ considering specific duration for inflation. For example, for $N=60$ we obtain $\alpha \geq 18.8$.

Then, we examined the observational predictions of the model for $n_{\rm s}$ and $r$ with respect to $\alpha$ parameter. It was found that, the scalar spectral index of this model shows degeneracy at large values of $\alpha$ parameter (see Fig. \ref{fig:ns}).
In the following, we examined three parameters of the reheating epoch, including the equation of state parameter $\omega_{\rm re}$, the duration of the reheating epoch $N_{\rm re}$, and the temperature of the reheating epoch $T_{\rm re}$.
We showed that the degeneracy of the model with respect to the parameter $\alpha$ is broken by analyzing  $N_{\rm re}$ and $T_{\rm re}$   as functions of $\alpha$ for fixed values of the $e$-folds number $N$.
The condition $ N_{\rm re}\geq 0 $ impose an upper bound on  the length of the inflation epoch as $N\leq55.7$. On the other side, from the $r-n_{\rm s}$ diagram we have $N\geq55$. Altogether, the condition $ N_{\rm re}\geq 0 $ thereto $r-n_{\rm s}$ observational data specified the allowed length of the inflationary period as $55\leq N\leq55.7$ and subsequently the non-canonical parameters $\alpha$ and $\mu$ were constrained (see Table \ref{tab1}). For $N=55$ and $N=55.7$, the allowed ranges of the $\alpha$ parameter are obtained as $110 \leq \alpha \leq 130$ and $60 \leq \alpha \leq 65$, respectively.

In addition, we examined the relic GWs for this model. We checked the present density of the relic GWs for different values of $\alpha$ parameter and duration of inflation between 55 to 55.7 $e$-folds.
We showed that the present density of relic GWs for this model,  falls within the sensitivity ranges of developing GWs detectors.

\subsection*{Acknowledgements}
The authors thank Soma Heydari and Milad Solbi for valuable discussions.

\subsection*{Data Availability Statement} No Data associated in the manuscript.



\begin{thebibliography}{0}
\expandafter\ifx\csname natexlab\endcsname\relax\def\natexlab#1{#1}\fi
\expandafter\ifx\csname bibnamefont\endcsname\relax
  \def\bibnamefont#1{#1}\fi
\expandafter\ifx\csname bibfnamefont\endcsname\relax
  \def\bibfnamefont#1{#1}\fi
\expandafter\ifx\csname citenamefont\endcsname\relax
  \def\citenamefont#1{#1}\fi
\expandafter\ifx\csname url\endcsname\relax
  \def\url#1{\texttt{#1}}\fi
\expandafter\ifx\csname urlprefix\endcsname\relax\def\urlprefix{URL }\fi
\providecommand{\bibinfo}[2]{#2}
\providecommand{\eprint}[2][]{\url{#2}}

\end{thebibliography}


\begin{thebibliography}{100}
\bibitem{Starobinsky:1980}
A.~A.~Starobinsky, \plb {\bf 91}, 99  (1980).
\bibitem{Guth:1981}
A. H. Guth, \prd {\bf 23}, 347 (1981).
\bibitem{Linde:1982}
A. D. Linde, \plb {\bf 108}, 389 (1982).
\bibitem{Albrecht:1982}
A.~Albrecht and P.~J.~Steinhardt, \prl {\bf 48}, 1220 (1982).
\bibitem{Linde:1990}
A.~D.~Linde, \textit{Particle Physics and Inflationary Cosmology}, Harwood, Chur, Switzerland (1990).
\bibitem{Baumann:2009}
D. Baumann, arXiv:0907.5424.
\bibitem{Liddle:2000}
A. R. Liddle and D. H. Lyth, \textit{Cosmological inflation and large scale structure}, Cambridge
University Press (2000).
\bibitem{BICEP2:2018}
P. A. R. Ade \textit{et al.} (Keck Array and bicep$2$ Collaborations), Phys. Rev. Lett. \textbf{121}, 221301 (2018).
\bibitem{Kallosh:2013a}
R. Kallosh and A. Linde, J. Cosmol. Astropart. Phys. \textbf{07}, 002 (2013).
\bibitem{Kallosh:2013}
R. Kallosh, A. Linde, and D. Roest, J. High. Energy. Phys. \textbf{11}, 198 (2013).

\bibitem{akrami:2020}
Y. Akrami \textit{et al.} (Planck Collaboration), Astron. Astrophys. {\bf 641}, A9 (2020).

\bibitem{Grishchuk:1974}
L. P. Grishchuk, Zh. Eksp. Teor. Fiz. \textbf{67}, 825 (1974).
\bibitem{Starobinsky:1979}
A. A. Starobinsky, \textit{JETP Lett}. {\bf 30}  682 (1979).
\bibitem{Starobinsky:1981}
A. A. Starobinsky, J. Exp. Theor. Phys. \textbf{34}, 438 (1981).

\bibitem{Kofman:1994}
L. Kofman, A. D. Linde, and A. A. Starobinsky, Phys. Rev. Lett. \textbf{73}, 3195 (1994).
\bibitem{Kofman:1996}
L. A. Kofman, arXiv:astro-ph/9605155.

\bibitem{Gialamas:2020}
I. D. Gialamas and A. B. Lahanas, Phys. Rev. D {\bf 101}, 084007 (2020).

\bibitem{Odintsov:2023}
S. D. Odintsov and T. Paul, Phys. of the Dark Universe {\bf 42}, 101263 (2023).


\bibitem{Martin:2010}
J. Martin and C. Ringeval,  Phys. Rev. D {\bf 82}, 023511 (2010).
\bibitem{LiangDai:2014}
L. Dai, M. Kamionkowski, and J. Wang, Phys. Rev. Lett. {\bf 113}, 041302 (2014).
\bibitem{Eshaghi:2016}
M. Eshaghi,  M. Zarei,  N. Riazi, and A. Kiasatpour, Phys. Rev. D {\bf 93}, 123517 (2016).
\bibitem{Mishra:2021}
S. S. Mishra, V. Sahnia, and A. A. Starobinsky, J. Cosmol. Astropart. Phys. {\bf 05}, 075 (2021).

\bibitem{Panotopoulos-2007}
G. Panotopoulos, Phys. Rev. D  {\bf 76}, 127302 (2007).
\bibitem{Unnikrishnan:2012}
S. Unnikrishnan, V. Sahni, and A. Toporensky, J. Cosmol. Astropart. Phys. {\bf 08}, 018 (2012).
\bibitem{Rezazadeh:2015}
K. Rezazadeh, K. Karami, and P. Karimi, J. Cosmol. Astropart. Phys. {\bf 09}, 053 (2015).

\bibitem{Mishra:2022}
S. S. Mishra and V. Sahni, arXiv:2202.03467.
\bibitem{Heydari:2024a}
S. Heydari and K. Karami, J. Cosmol. Astropart. Phys. {\bf 02}, 047 (2024).
\bibitem{Heydari:2024b}
S. Heydari and K. Karami, Eur. Phys. J. C  {\bf 84}, 127 (2024).
\bibitem{Heydari:2024c}
S. Heydari and K. Karami, Astrophys. J. {\bf 975}, 148 (2024).

\bibitem{Garriga:1999}
J. Garriga and  V. F. Mukhanov, Phys. Lett. B {\bf 458}, 219 (1999).

\bibitem{akrami:2018}
Y. Akrami \textit{et al.} (Planck Collaboration), Astron. Astrophys. {\bf 641}, A10 (2020).
\bibitem{bk18}
P. A. R. Ade \textit{et al.} (BICEP/Keck Collaboration), Phys. Rev. Lett. {\bf 127}, 151301 (2021).
\bibitem{Paoletti:2022}
D. Paoletti, F. Finell, J. Valiviita, and M. Hazumi, Phys. Rev. D {\bf 106}, 083528 (2022).


\bibitem{Tanabashi:2018}
M. Tanabashi \textit{et al.}, Phys. Rev. D {\bf 98}, 030001 (2018).
\bibitem{Workman:2022}
R. L. Workman \textit{et al.} (Particle Data Group), Prog. Theor. Exp. Phys. {\bf 2022}, 083C01 (2022).
\bibitem{Chen:2007}
X. Chen, M. X. Huang, S. Kachru, and G. Shiu, J. Cosmol. Astropart. Phys. {\bf 01}, 002 (2007).
%

\bibitem{Martin2015} J. Martin, C. Ringeval, and V. Vennin, Phys. Rev. Lett. {\bf 114}, 081303 (2015).
\bibitem{Munoz2015} J. B. Muñoz and M. Kamionkowski, Phys. Rev. D {\bf 91}, 043521 (2015).
\bibitem{Cook2015} J. L. Cook, E. Dimastrogiovanni, D. A. Easson, and L. M. Krauss, J. Cosmol. Astropart. Phys. {\bf 04}, 047 (2015).

\bibitem{German2020} G. German, J. Cosmol. Astropart. Phys. {\bf 11}, 006 (2020).






\bibitem{allen88}
B.~Allen, Phys. Rev. D {\bf 37}, 2078 (1988).
\bibitem{sahni:1990}
V.~Sahni, Phys. Rev. D {\bf 42}, 453 (1990).

\bibitem{sami2002}
V. Sahni, M. Sami, and T. Souradeep, Phys. Rev. D {\bf 65}, 023518 (2002).

\bibitem{dany_18}
C.~Caprini and D.~G.~Figueroa, Class. Quant. Grav. {\bf 35}, 163001 (2018).
\bibitem{dany_19}
D.~G.~Figueroa and E.~H.~Tanin,  J. Cosmol. Astropart. Phys. {\bf 08}, 011 (2019).
\bibitem{Bernal:2019lpc}
N.~Bernal and F.~Hajkarim, Phys. Rev. D {\bf 100}, 063502 (2019).


%
%
%
%




\bibitem{Crowder:2005BBO}
J. Crowder, and N. J. Cornish, Phys. Rev. D {\bf 72}, 083005 (2005).
\bibitem{Corbin:2006BBO}
V. Corbin and N. J. Cornish, Class. Quant. Grav. {\bf 23}, 2435 (2006).
\bibitem{Harry:2006BBO}
G. M. Harry et al., Class. Quant. Grav. {\bf 23}, 4887 (2006).
\bibitem{Yagi:2011BBODECIGO}
K. Yagi and N. Seto, Phys. Rev. D {\bf 83}, 044011 (2011).
\bibitem{Yagi:2017BBODECIGO}
K. Yagi and N. Seto, Phys. Rev. D {\bf 95}, 109901 (2017).
\bibitem{Seto:2001DECIGO}
N. Seto, S. Kawamura, and T. Nakamura, Phys. Rev. Lett. {\bf 87}, 221103 (2001).
\bibitem{Kawamura:2006DECIGO}
S. Kawamura \textit{et al.}, Class. Quant. Grav. {\bf 23}, S125 (2006).
\bibitem{Kawamura:2011DECIGO}
S. Kawamura \textit{et al.}, Class. Quant. Grav. {\bf 28}, 094011 (2011).
%




\bibitem{skaCarilli:2004}
C. L. Carilli, and S. Rawlings, New Astron. Rev. {\bf 48}, 979 (2004).
\bibitem{ska}
C. J. Moore, R. H. Cole, and C. P. L. Berry, Class. Quant. Grav. {\bf 32}, 015014 (2015).
\bibitem{skaWeltman:2020}
A. Weltman \textit{et al.}, Publ. Astron. Soc. Aust. {\bf 37}, e002 (2020).
%
%
%

\end{thebibliography}
\end{document}